\documentclass[aps, prd, onecolumn, amsmath, superscriptaddress, floatfix, nofootinbib, preprintnumbers, preprint]{revtex4-1}

\usepackage{mathrsfs}
\usepackage{amsfonts}
\usepackage{amsmath, amssymb}
\usepackage{physics}
\usepackage{array}
\usepackage{verbatim}
\usepackage{epsfig}
\usepackage{graphicx}
\usepackage[colorlinks=true,citecolor=blue,linkcolor=blue,urlcolor=blue]{hyperref}
\usepackage[normalem]{ulem}
\usepackage{xcolor}
\usepackage{multirow}
\usepackage[caption=false]{subfig}

\usepackage{graphicx}
\graphicspath{{./figures/}}
\usepackage{dcolumn}
\usepackage{bm}
\usepackage{slashed}
\usepackage{braket}
\usepackage{siunitx}
\usepackage{xpatch}
\usepackage[mathlines]{lineno}
\usepackage{comment}
\usepackage{soul}
\usepackage{xcolor}
\usepackage{xfrac}

\newcommand{\diff}[1]{\mathrm{d}#1}

\def\bea#1\eea{\begin{align}#1\end{align}}

\newcommand{\Ph}{P_h}
\newcommand{\PhT}{P_{hT}}
\newcommand{\Mh}{M_h}
\newcommand{\MhT}{M_{hT}}
\newcommand{\sigh}{\sigma_h}

\newcommand{\eref}[1]{Eq.~(\ref{eq:#1})}

\newcommand{\fref}[1]{Fig.~\ref{fig:#1}}

\newcommand{\sref}[1]{Sec.~\ref{sec:#1}}
\newcommand{\ssref}[1]{Sec.~\ref{ss:#1}}

\begin{document}

\title{Accessing gluon polarization with high-$P_T$ hadrons in SIDIS}

\author{R. M. Whitehill}
\affiliation{Department of Mathematics, Statistics, and Physics,
        Wichita State University, Wichita, Kansas 67260, USA}
\author{Yiyu Zhou}
\affiliation{Guangdong Provincial Key Laboratory of Nuclear Science, 
        Institute of Quantum Matter,
        South China Normal University, Guangzhou 510006, China}
\affiliation{Guangdong-Hong Kong Joint Laboratory of Quantum Matter,
        Southern Nuclear \\ Science Computing Center, 
        South China Normal University, Guangzhou 510006, China}
\affiliation{Department of Physics and Astronomy, University of California,
        Los Angeles, California 90095, USA}
\author{N. Sato}
\affiliation{Jefferson Lab, Newport News, Virginia 23606, USA}
\author{W. Melnitchouk}
\affiliation{Jefferson Lab, Newport News, Virginia 23606, USA}
\affiliation{\mbox{CSSM and CDMPP, Department of Physics, University of Adelaide, Adelaide 5000, Australia} \\
        \vspace*{0.2cm}
        {\bf Jefferson Lab Angular Momentum (JAM) Collaboration
        \vspace*{0.2cm}}}

\date{\today}
\preprint{JLAB-THY-22-3743, ADP-22-30/T1201}

\begin{abstract}
A recent global QCD analysis of jet production and other polarized scattering data has found the presence of negative solutions for the gluon helicity distribution in the proton, $\Delta g$, along with the traditional $\Delta g > 0$ solutions.
We consider polarized semi-inclusive deep-inelastic scattering for hadrons produced with large transverse momentum as a means of constraining the dependence of $\Delta g$ on the parton momentum fraction, $x$.
Focusing on the double longitudinal spin asymmetry, we identify the kinematics relevant for future experiments at Jefferson Lab and the Electron-Ion Collider which are particularly sensitive to the polarized gluon channel and could discriminate between the different $\Delta g$ behaviors.
We find that a $\sim 20$~GeV beam at the high luminosity Jefferson Lab may be especially well-suited for discriminating between the positive and negative solutions.
\end{abstract}

\maketitle

\section{Introduction}

Understanding the proton spin puzzle -- the detailed decomposition of the proton's spin into its quark and gluon helicity and orbital angular momentum components -- has been one of the major drivers of new experimental programs in the nuclear physics community over the past three decades~\cite{Aidala:2012mv}.
Worldwide efforts at various accelerator facilities have now established fairly comprehensively the total fraction of the helicity carried by quarks, although some questions still remain about the detailed flavor decomposition of the sea quark contributions.
A major breakthrough was made with the observation of significant double spin asymmetries in inclusive jet production in polarized proton--proton collisions at RHIC~\cite{STAR:2014wox}, from which the first clearly nonzero signal for a polarized gluon distribution was extracted~\cite{deFlorian:2014yva}.
Subsequent inclusive jet production data from the STAR~\cite{STAR:2019yqm, STAR:2021mfd, STAR:2021mqa} and PHENIX~\cite{PHENIX:2010aru} collaborations have reaffirmed these observations, leading to a greater confidence that finally both the quark and gluon helicity content of the proton may be relatively well understood.

\begin{figure}[t]
\centering
\includegraphics[width=0.65\textwidth]{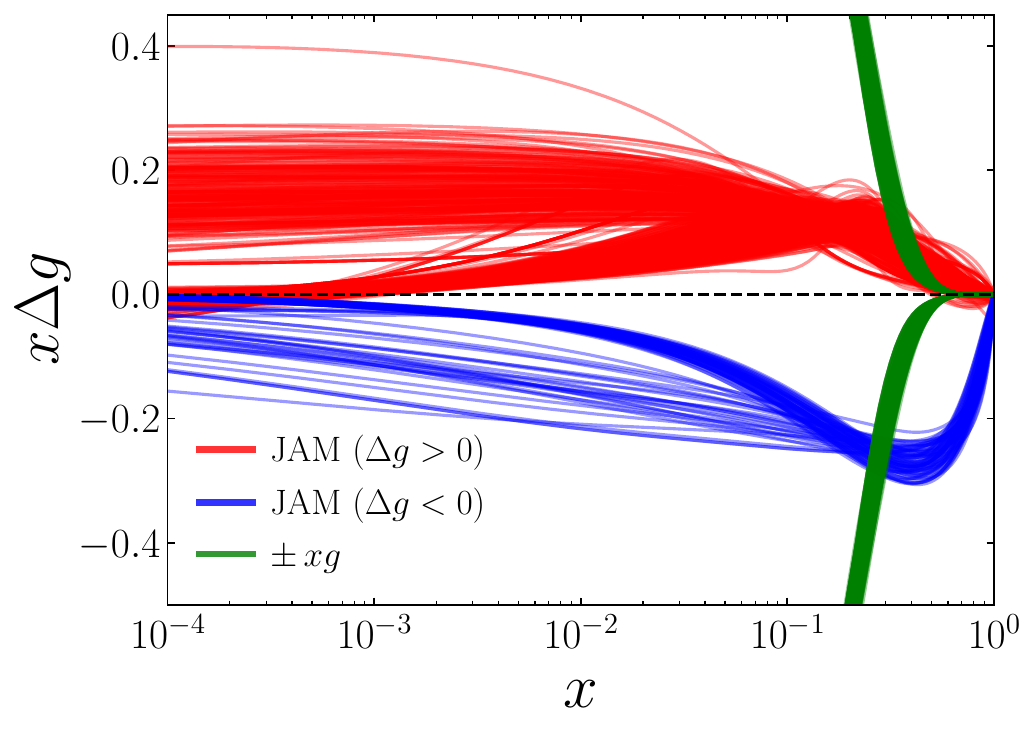}
\caption{Polarized gluon distribution $x \Delta g(x)$ at $Q^2 = 10$~GeV$^2$ from the recent JAM global QCD analysis~\cite{Zhou:2022wzm}, showing separately solutions with $\Delta g > 0$ (red lines) and $\Delta g < 0$ (blue lines), and compared with $\pm$ the unpolarized gluon distribution, $xg(x)$ (green lines).}
\label{fig:gluon-ppdfs}
\end{figure}

Recently, the JAM Collaboration~\cite{Zhou:2022wzm} revisited the analysis of the jet data to examine the extent to which these results depend on the theoretical assumptions made in the analysis, such as SU(3) flavor symmetry for the axial vector charges that constrain nonsinglet combinations of spin-dependent parton distribution functions (PDFs)~\cite{Bass:2009ed, Ethier:2017zbq}, and positivity constraints for unpolarized PDFs~\cite{Candido:2020yat, Collins:2021vke}.
In particular, the analysis found that without the PDF positivity constraints, which are not formally required on theoretical grounds~\cite{Collins:2021vke}, a second set of solutions is possible for which $\Delta g < 0$, as illustrated in Fig.~\ref{fig:gluon-ppdfs}.
This set of solutions is permitted because the double spin asymmetry for inclusive jet production is quadratic in the parton polarization.
That is, the traditional small and positive $\Delta g$ and the positive quark polarization $\Delta q$ combine to produce an overall positive asymmetry, as observed in the STAR data.
However, the data also allow negative $\Delta g$ with larger magnitude, which combines with the positive $\Delta q$ to produce a cancellation between a positive contribution from the gluon-gluon channel and a negative contribution from the quark-gluon channel, giving equally good descriptions of the inclusive jet data.

\begin{figure}[t]
\centering
\includegraphics[width=0.75\textwidth]{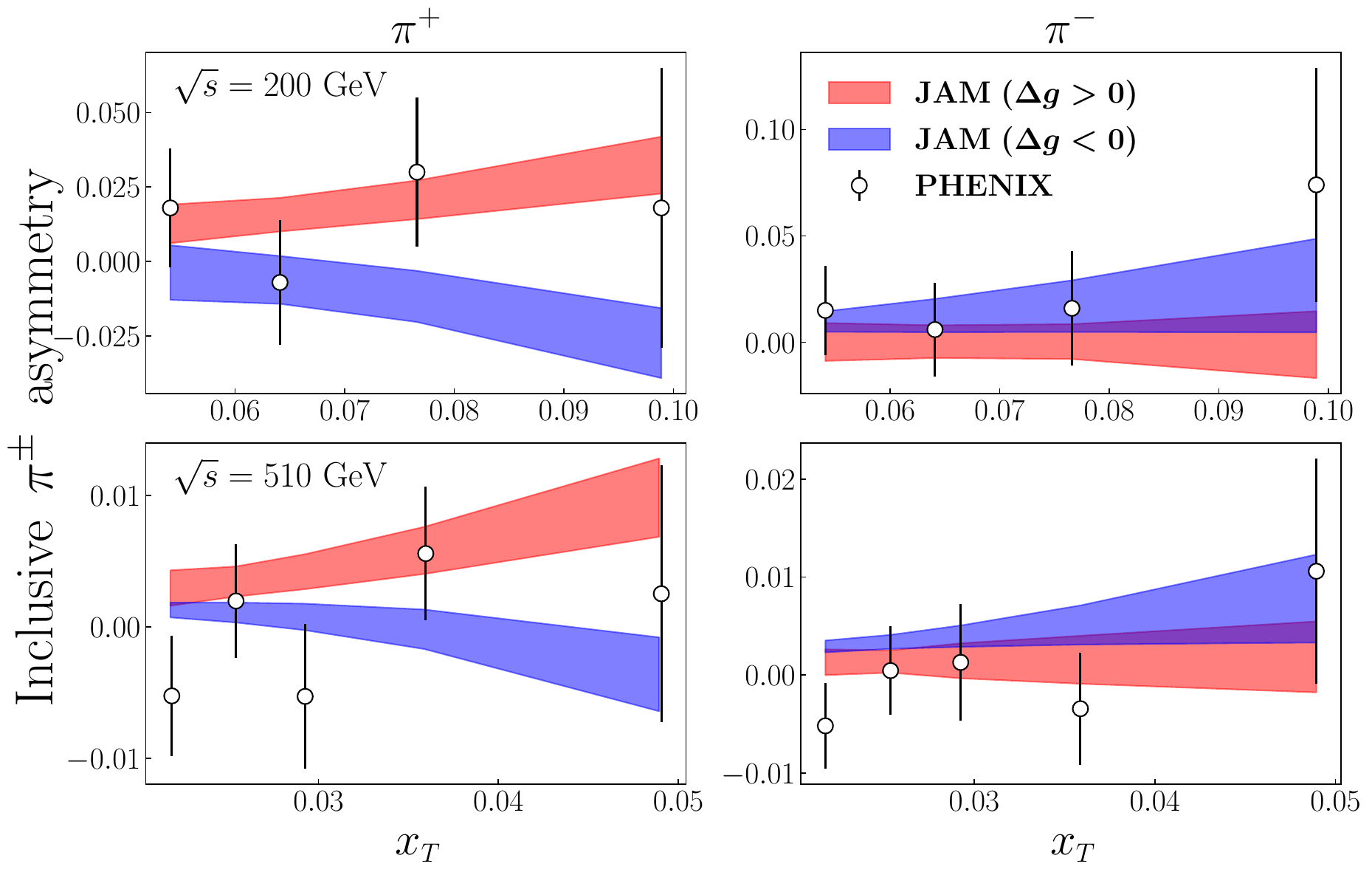}
\caption{Inclusive double spin asymmetry for $\pi^+$ (left column) and $\pi^-$ (right column) production from the PHENIX experiment~\cite{PHENIX:2014axc, PHENIX:2020trf} at $\sqrt{s}=200$~GeV (top row) and 510~GeV (bottom row) as a function of $x_T=2p_T/\sqrt{s}$, compared with the predictions from the JAM analysis~\cite{Zhou:2022wzm} with positive (red bands) and negative (blue bands) gluon helicity.}
\label{fig:RHIC}
\end{figure}

Earlier, J\"{a}ger {\it et al.}~\cite{Jager:2003ch} discussed possible constraints on the sign of $\Delta g$ from inclusive pion production in polarized $pp$ collisions.
In particular, they considered PHENIX data~\cite{PHENIX:2015fxo} for neutral pions produced with relatively low transverse momentum, $1 < p_T < 5$~GeV, within a collinear factorization framework.
While the applicability of collinear factorization in this region may be questionable, J\"{a}ger {\it et al.} extracted a small but negative lower bound for the double spin asymmetry by finding the extremum of the asymmetry in Mellin space as a function of the moments of $\Delta g$.
They found that the extremum corresponds to a negative gluon helicity, with associated violations of PDF positivity bounds at higher $x$ values, similar to those found in the JAM analysis~\cite{Zhou:2022wzm}.

Moreover, the sign of $\Delta g$ was investigated by comparing PHENIX data on inclusive charged pion production~\cite{PHENIX:2014axc, PHENIX:2020trf}.
A positive $\Delta g$ was expected to produce a hierarchy of double spin asymmetries with that for $\pi^+ > \pi^0 > \pi^-$.
In Fig.~\ref{fig:RHIC} we show the $\pi^+$ and $\pi^-$ asymmetries at $pp$ center of mass energies $\sqrt{s} = 200$ and 510~GeV as a function of $x_T = 2 p_T/\sqrt{s}$, where $p_T$ is the transverse momentum of the final state pion in the laboratory frame, compared with predictions from the recent JAM analysis~\cite{Zhou:2022wzm}.
While the $\pi^+$ asymmetry in particular has the potential to discriminate between the different $\Delta g$ solutions, the uncertainties on these data do not exclude either a positive or negative $\Delta g$.

A possible way to resolve this problem would be to identify observables that are linear in the gluon polarization $\Delta g$, and where the gluon contribution is not suppressed relative to the quark contribution, as it is in inclusive deep-inelastic scattering (DIS).
One candidate process is polarized lepton-nucleon semi-inclusive DIS (SIDIS), with production of hadrons in the final state with large transverse momentum.
Here, the contribution in which the hard scattering involves an initial state gluon enters at the same order in the strong coupling $\alpha_s$ as the quark scattering contribution.
One therefore expects greater sensitivity to the $\Delta g$ PDF at high transverse momentum than at low transverse momentum.

In this paper, we examine the polarized SIDIS process for the production of charged pions at large transverse momentum, with the aim of assessing its potential impact on $\Delta g$.
In \sref{kinematic-definitions-and-collinear-factorization} we outline the theoretical framework used in the analysis, including cross section definitions and the results within collinear factorization in terms of PDFs and fragmentation functions.
The perturbative calculation of the hard scattering amplitudes is presented in \sref{hard-scattering-amplitudes} for both spin-averaged and spin-dependent scattering.
In \sref{results} we discuss the results of our numerical simulations at kinematics relevant for Jefferson Lab with the current 12~GeV and a possible future energy upgraded 22~GeV electron beam, as well as at the EIC, with emphasis on the efficacy of discriminating the two $\Delta g$ solutions in \fref{gluon-ppdfs}.
Finally, in Sec.~\ref{sec:conc} we summarize our findings and discuss possible future extensions of this work.

\section{Theoretical framework}
\label{sec:kinematic-definitions-and-collinear-factorization}

In this section we present the theoretical framework which we use to compute the relevant cross sections in this analysis.
After introducing the basic definitions of the observables and their kinematics, we then specialize to the case where these are computed using the collinear factorization approximation.

\subsection{Definitions and kinematics}
\label{ssec:definitions}

The process that we consider in this work is the leptoproduction of a charged hadron $h$ in SIDIS from a nucleon $N$,
\begin{equation}
    e^{}\, (\ell) + N (P) \to e^{}\, (\ell') + h (\Ph) + X,
\end{equation}
where $\ell$ and $\ell'$ are the incoming and scattered lepton four-momenta, $P$ and $\Ph$ are target nucleon and produced hadron four-momenta, respectively, and $X$ denotes unobserved final state particles.
The scattering cross section for this reaction can be written as a tensor product of a leptonic tensor $L_{\mu \nu}$ and a hadronic tensor $W^{\mu \nu}$,
\begin{eqnarray}
\label{eq:gen.xsec}
    4 \Ph^0 E'\frac{\diff\sigh}{\diff^3\bm{\ell'} \, \diff^3\bm{\Ph}}
    = \frac{2\alpha^2}{sQ^4}\, L_{\mu\nu}\, W^{\mu\nu},
\end{eqnarray}
where $\alpha = e^2/4\pi$ is the electromagnetic coupling, $E'$ and $\Ph^0$ are the final state lepton and hadron energies, respectively, 
    $Q^2 = -(\ell - \ell')^2$ 
is the squared four-momentum transfer to the nucleon, and 
    $s = (\ell + P)^2$
is the squared center of mass energy.
In terms of Lorentz invariant variables, after integrating over azimuthal angles, \eref{gen.xsec} can also be written as
\begin{eqnarray}
\label{eq:upol.xsec}
    \frac{\diff\sigh}{\diff x \, \diff y \, \diff z \, \diff \PhT^2} 
    = \frac{\pi^2\alpha^2y}{2zQ^4}\, L_{\mu\nu}\, W^{\mu\nu}, 
\end{eqnarray}
where the transverse momentum of the produced hadron $\bm{P}_{hT}$ is defined in the ``photon frame'', in which the incoming nucleon and photon are back-to-back.
The Bjorken scaling variable $x$, lepton inelasticity $y$, and fragmentation variable $z$ in Eq.~(\ref{eq:upol.xsec}) are defined as
\begin{eqnarray}
\label{eq:scal.var}
    x = \frac{Q^2}{2P \cdot q}, \qquad 
    y = \frac{P \cdot q}{P \cdot \ell}, \qquad 
    z = \frac{P \cdot \Ph}{P \cdot q}, 
\end{eqnarray}
respectively, where $q = \ell - \ell'$ is the four-momentum transfer.
Since we consider the polarizations for the initial lepton and nucleon, both the leptonic and hadronic tensors can in general be decomposed into helicity-independent parts, which are symmetric in the indices $\{ \mu, \nu \}$, and helicity-dependent parts, which are antisymmetric,
\begin{subequations}
\begin{eqnarray}
\label{eq:tensor.decomp}
    L_{\mu \nu} &=& L_{\mu \nu}^{S} + \lambda_{\ell}\, L_{\mu \nu}^{A}, \\
    W^{\mu \nu} &=& W^{\mu \nu}_{S} + \lambda\, W^{\mu \nu}_{A},
\end{eqnarray}
\end{subequations}
where $\lambda_{\ell}$ and $\lambda$ are the helicities of the incident lepton and nucleon.
As the products between symmetric and antisymmetric terms vanish, we can write the tensor product as a sum of purely symmetric and purely antisymmetric components,
\begin{eqnarray}
\label{eq:simplified.tens.prod}
    L_{\mu \nu} W^{\mu\nu} = L_{\mu \nu}^{S} W^{\mu \nu}_{S} 
    + \lambda_{\ell}\, \lambda\, L_{\mu \nu}^{A} W^{\mu \nu}_{A}.
\end{eqnarray}

The unpolarized cross section is calculated by averaging over the four possible combinations of the lepton ($\rightarrow$) and nucleon ($\Rightarrow$) helicities,
\begin{subequations}
\label{eq:upolandpol.diff.xsec}
\begin{eqnarray}
\label{eq:upol.diff.xsec}
    \diff\sigma_h = \frac14
    \left[
    \diff\sigh^{\stackrel{\rightarrow}{\Rightarrow}} +
    \diff\sigh^{\stackrel{\leftarrow}{\Leftarrow}} +
    \diff\sigh^{\stackrel{\rightarrow}{\Leftarrow}} + \diff\sigh^{\stackrel{\leftarrow}{\Rightarrow}} 
    \right],
\end{eqnarray}
leaving only the product from the contraction of the symmetric parts of the leptonic and hadronic tensors.
For the polarized cross section, one takes the difference between the aligned and antialigned spin states,
\begin{eqnarray}
\label{eq:pol.diff.xsec}
    \diff\Delta\sigh = \frac{1}{4}\left[
    \diff\sigh^{\stackrel{\rightarrow}{\Rightarrow}} +
    \diff\sigh^{\stackrel{\leftarrow}{\Leftarrow}} -
    \diff\sigh^{\stackrel{\rightarrow}{\Leftarrow}} - \diff\sigh^{\stackrel{\leftarrow}{\Rightarrow}} 
    \right],
\end{eqnarray}
\end{subequations}
which gives the contribution from the antisymmetric parts of the tensors.

Since the helicity dependent terms are proportional to
    $\lambda_\ell \lambda$,
one has
    $\diff\sigh^{\stackrel{\rightarrow}{\Rightarrow}}
    = \diff\sigh^{\stackrel{\leftarrow}{\Leftarrow}}$
and
    $\diff\sigh^{\stackrel{\rightarrow}{\Leftarrow}}
    = \diff\sigh^{\stackrel{\leftarrow}{\Rightarrow}}$,
so that only the relative handedness of the lepton and nucleon are relevant.
Fixing the initial nucleon to be right-handed, 
    $\lambda = 1$,
we can simplify Eqs.~(\ref{eq:upolandpol.diff.xsec}) to write the cross sections in terms of the two possible relative alignments,
\begin{subequations}
\begin{eqnarray}
\label{eq:upol.diff.xsec.simplified}
    \diff\sigh = \frac{1}{2}\left[
    \diff\sigh^{\stackrel{\rightarrow}{\Rightarrow}} + \diff\sigh^{\stackrel{\leftarrow}{\Rightarrow}} 
    \right],
\end{eqnarray}
and
\begin{eqnarray}
\label{eq:pol.diff.xsec.simplified}
    \diff\Delta\sigh = \frac{1}{2}\left[
    \diff\sigh^{\stackrel{\rightarrow}{\Rightarrow}} - \diff\sigh^{\stackrel{\leftarrow}{\Rightarrow}} 
    \right].
\end{eqnarray}
\end{subequations}
In the one-photon exchange approximation, the lepton tensor is given by
\begin{eqnarray}
\label{eq:lep.tens}
    L_{\mu\nu} = 2(\ell_{\mu}\ell'_{\nu} + \ell'_{\mu}\ell_{\mu} - g_{\mu\nu}\ell \cdot \ell' + i\epsilon_{\mu\nu\alpha\beta}s^{\alpha}q^{\beta}).
\end{eqnarray}
Since the lepton is longitudinally polarized with helicity
    $\lambda_\ell$,
the lepton spin vector can be written as
    $s^{\mu} = \lambda_\ell\, \ell^{\mu}$,
and \eref{lep.tens} becomes
\begin{eqnarray}
\label{eq:hel.lep.tens}
    L_{\mu\nu} = 2(\ell_{\mu}\ell'_{\nu} + \ell'_{\mu}\ell_{\mu} - g_{\mu\nu}\ell \cdot \ell' - i\lambda_\ell\, \epsilon_{\mu\nu\alpha\beta}\ell^{\alpha}\ell'^{\beta}).
\end{eqnarray}
Note that we assume a kinematic regime where
    $x$
and 
    $1/Q$
are both small enough that we can neglect the lepton and parton masses.
However, in this study we keep the nucleon and final state hadron masses, since terms proportional to 
    $M/Q$ 
and
    $\Mh/Q$
in the calculations are not necessarily insignificant relative to the massless limit at the lower energy scales of Electron-Ion Collider (EIC) and Jefferson Lab kinematics.
We do not consider other $1/Q$ corrections, such as those associated with dynamical higher twists effects involving multiparton correlations, which would require a more detailed analysis than that undertaken in this study.

The hadronic tensor for SIDIS can be written as
\begin{eqnarray}
\label{eq:gen.had.tens}
    W^{\mu\nu} = \frac{1}{(2\pi)^4}\sum_{X}\int \diff^4z\, e^{iq \cdot z} \bra{P,S}J^{\mu}(z)\ket{\Ph,X}\bra{\Ph,X}J^{\nu}(0)\ket{P,S},
\end{eqnarray}
where $J^\mu$ is the electromagnetic current.
In this work we will focus on the large transverse momentum region by using collinear factorization, following a similar procedure as in Ref.~\cite{Wang:2019bvb}. 
Since our aim is to compute the total asymmetry after integration of the angular modulations between the lepton plane and the hadronic plane, it will not be necessary to decompose the hadronic tensor into all possible angular modulations.

\subsection{Collinear factorization}

The expressions for the cross sections simplify considerably at large $Q^2$ and large $\PhT$, where they can be factorized in terms of partonic subprocesses.
In particular, using the factorization formalism from Wang {\it et al.}~\cite{Wang:2019bvb}, the hadronic cross sections can be written in terms of partonic cross sections and the corresponding nonperturbative functions,
\begin{subequations}
\label{eq:fact.thm}
\begin{eqnarray}
    4 \Ph^0 E'\frac{\diff\sigh}{\diff^3\bm{\ell}'\diff^3\bm{P}_h}
    &=& \sum_{ij} \int_{x}^{1}\frac{\diff\xi}{\xi}\int_{z}^{1}\frac{\diff\zeta}{\zeta^2} \left(4 k_1^0 E'\frac{\diff\hat{\sigma}_{ij}}{\diff^3\bm{\ell}'\diff^3\bm{k}_1}\right) f_{i/N}(\xi)\, D_{h/j}(\zeta), 
\\
&& \hspace*{-5.1cm} {\rm and}
\notag
\\
    4 \Ph^0 E'\frac{\diff\Delta\sigma_h}{\diff^3\bm{\ell}'\diff^3\bm{P}_h}
    &=& \sum_{ij} \int_{x}^{1}\frac{\diff\xi}{\xi}\int_{z}^{1}\frac{\diff\zeta}{\zeta^2} \left(4 k_1^0 E'\frac{\diff\Delta\hat{\sigma}_{ij}}{\diff^3\bm{\ell}'\diff^3\bm{k}_1}\right)\Delta f_{i/N}(\xi)\, D_{h/j}(\zeta),
\end{eqnarray}
\end{subequations}
for unpolarized and polarized scattering, respectively, where 
    $f_{i/N}$ and $\Delta f_{i/N}$ 
are the spin-averaged and spin-dependent PDFs of flavor $i$ in the nucleon $N$, and
    $D_{h/j}$
is the fragmentation function for a parton of flavor $j$ to hadronize to the hadron $h$.
The momentum fractions $\xi$ and $\zeta$ are defined in terms of the incident parton momentum $p$ and outgoing parton momentum $k_1$ by
\begin{eqnarray}
\label{eq:part.mom}
    p = \xi P, \qquad 
    k_1 = \frac{\Ph}{\zeta}.
\end{eqnarray}

In analogy with the hadronic cross section, the partonic cross section can be expressed in terms of the partonic structure tensor
    $\widehat{W}_{ij}^{\mu\nu}$,
which describes the interactions in the hard scattering process at the parton level,
\begin{eqnarray}
\label{eq:part.xsec}
    4k_1^0E'\frac{\diff\hat{\sigma}_{ij}}{\diff^3\bm{\ell}'\diff^3\bm{k}_1}
    = \frac{2\alpha^2}{\hat{s}Q^4}L_{\mu\nu}\widehat{W}_{ij}^{\mu\nu}.
\end{eqnarray}
Again, in analogy with the hadronic tensor, the partonic structure tensor is defined as
\begin{eqnarray}
\label{eq:gen.part.tens}
    \widehat{W}^{\mu\nu}_{ij} 
    = \frac{1}{(2\pi)^4}\sum_{X}\int \diff^4w\, e^{iq \cdot w} 
    \bra{p_i,s} J^{\mu}(w) \ket{k_{1j},X} \bra{k_{1j},X} J^{\nu}(0) \ket{p_i,s}.
\end{eqnarray}
With the factorization theorem in Eqs.~(\ref{eq:fact.thm}), for the spin-averaged case we can relate the (symmetric) hadronic and partonic tensors by
\begin{eqnarray}
\label{eq:had-tens.fact}
    W_S^{\mu\nu} 
    = \sum_{ij} \int_{x}^{1}\frac{\diff\xi}{\xi} \int_{z}^{1}\frac{\diff\zeta}{\zeta^2}\ 
    \widehat{W}_{ij}^{\mu\nu}\, f_{i/P}(\xi)\, D_{h/j}(\zeta),
\end{eqnarray}
and similarly for the (antisymmetric) spin-dependent tensors.
Since the hadronic quantities are calculated from partonic quantities, it is useful to introduce the partonic variables
\begin{eqnarray}
\label{eq:hat.var}
    \hat{x} \equiv \frac{Q^2}{2p \cdot q} = \frac{x}{\xi},
    \qquad
    \hat{z} \equiv \frac{p \cdot k_1}{p \cdot q} = \frac{z}{\zeta},
    \qquad
    k_{1T} \equiv \frac{\PhT}{\zeta}.
\end{eqnarray}
Additionally, it will be convenient to define the scaled transverse momentum variable
\begin{equation}
\label{eq:qT.from.PT}
    q_{T} = \frac{\PhT}{z},
\end{equation}
which provides a scale separation relative to $Q$, so that in the regime where $q_T/Q \sim 1$ one can analyze the reaction using the collinear framework.

\section{Hard scattering amplitudes}
\label{sec:hard-scattering-amplitudes}

\begin{figure}[t]
\centering
\includegraphics[width=0.75\textwidth]{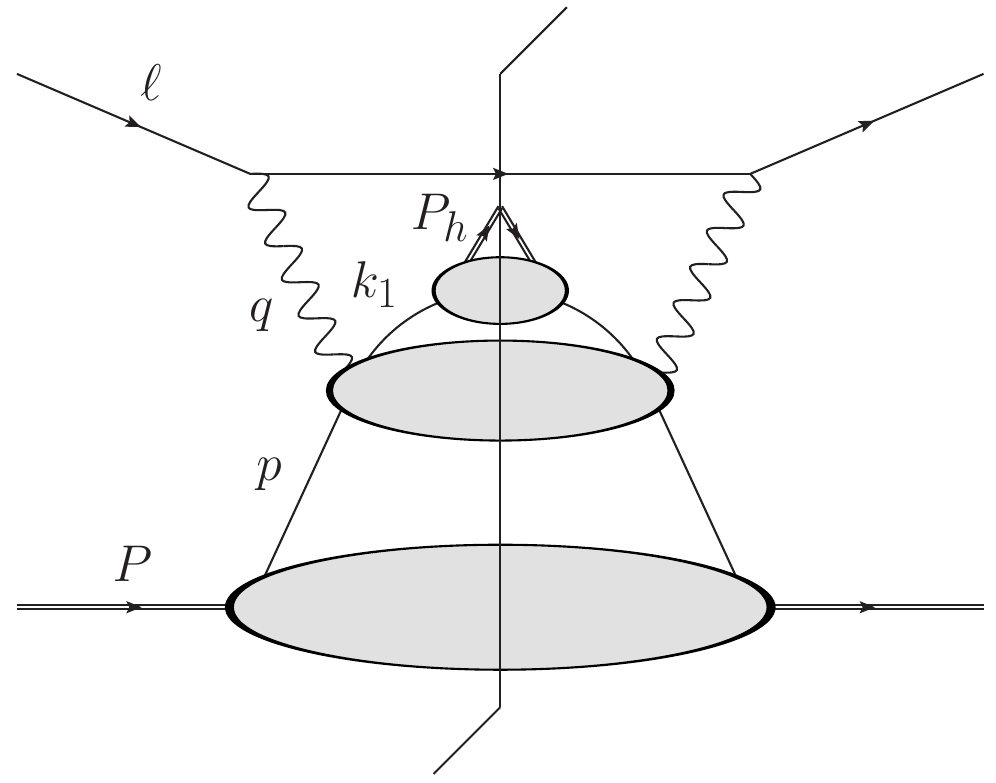}
\caption{Sketch of the diagram representing SIDIS of a lepton (momentum $\ell$) from a proton ($P$) in the one-photon ($q$) exchange approximation, with $p$ the momentum of the initial parton in the proton and $k_1$ the momentum of the parton fragmenting to the hadron $h$ (with momentum $P_h$). The lower blob represents the soft part of the scattering process, the central blob represents the hard part of the scattering process, and the upper blob represents the parton fragmentation to the produced hadron $h$.}
\label{fig:2-2.scat}
\end{figure}

The factorized form of the relations in Eqs.~(\ref{eq:fact.thm}) allows us to compute the SIDIS cross section from the parton level in terms of the partonic structure tensor (\ref{eq:gen.part.tens}) and the appropriate PDFs and fragmentation functions.
In this study, the partonic structure tensor, including phase space constraints,
\begin{eqnarray}
\label{eq:part-tens.amps}
    \widehat{W}_{ij}^{\mu\nu} = \int \diff \Pi\, \mathcal{M}^{\mu\dagger}_{i}\mathcal{M}^{\nu}_{j},
\end{eqnarray}
is calculated directly at leading order (LO) in the strong coupling for the 
    $2 \rightarrow 2$
scattering process.
A sketch of the SIDIS process is illustrated in Fig.~\ref{fig:2-2.scat}, and the four tree-level diagrams contributing to the partonic structure tensor are shown in \fref{Feyn.Diag}.
Note that the incoming photon and parton momenta are denoted as 
    $q$
and
    $p$,
respectively, the fragmenting parton has momentum 
    $k_1$,
while the other outgoing parton has momentum 
    $k_2$.

From momentum conservation, the outgoing two particle phase space is represented by
\begin{eqnarray}
\label{ps_space_factor}
    \diff \Pi = \frac{\diff^3 k_2}{(2\pi)^3 2 k_2^{0}}\,
    (2\pi)^4 \delta^{(4)}(q+p-k_1-k_2).
\end{eqnarray}
Integrating over $k_2$, we have
\begin{eqnarray}
\label{eq:simplified_ps_factor}
    \int \diff \Pi 
    = 2\pi\delta_{+}(k_2^2) 
    = \frac{2\pi \hat{x}}{Q^2}\,
    \delta \bigg((1-\hat{x})(1-\hat{z}) - \frac{\hat{x}\hat{z}q_{T}^2}{Q^2} \bigg),
\end{eqnarray}
where the $\delta$ function imposes a strict relationship between $\zeta$ and $\xi$ that depends on the hadronic kinematics, and the subscript ``+'' denotes the positive energy solution for $k_2^0$.
Solving \eref{simplified_ps_factor}, we find
\begin{eqnarray}
\label{eq:zeta-xi}
    \zeta = z\bigg[ 1 + \frac{x}{\xi - x}\, \frac{q_T^2}{Q^2} \bigg].
\end{eqnarray}
Noting that $\zeta$ monotonically decreases as a function of $\xi$, the lower bound for the factorization integral in $\xi$ is given by
\begin{eqnarray}
\label{eq:xi-min}
    \xi_{\rm min} = x \left[ 1 + \frac{z}{1-z}\frac{q_T^2}{Q^2} \right].
\end{eqnarray}

\begin{figure}
\centering
\quad
\subfloat[]{
\includegraphics[width = 0.32 \textwidth]{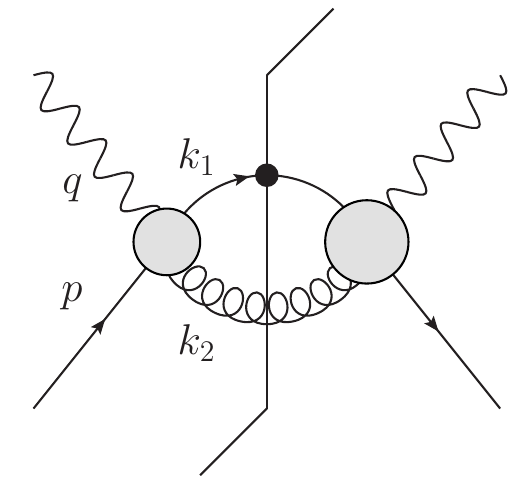}
\label{fig:FeynA}
}
\hfil
\subfloat[]{
\includegraphics[width = 0.32 \textwidth]{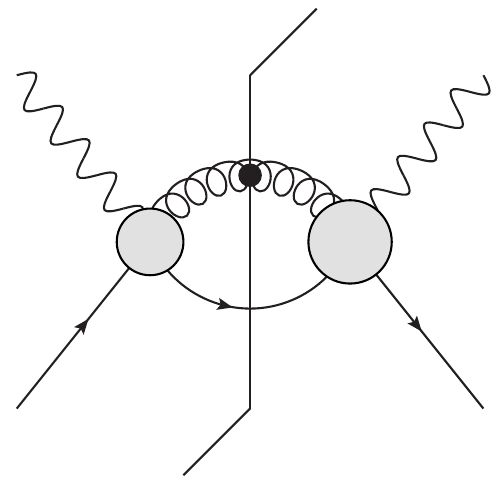}
\label{fig:FeynB}
}
\quad
\\
\quad
\subfloat[]{
\includegraphics[width = 0.32 \textwidth]{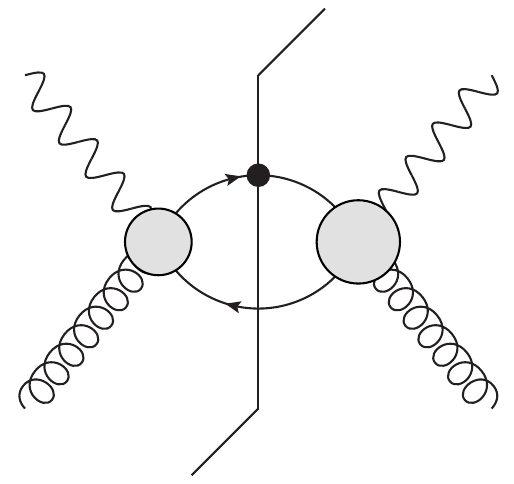}
\label{fig:FeynC}
}
\hfil
\subfloat[]{
\includegraphics[width = 0.32 \textwidth]{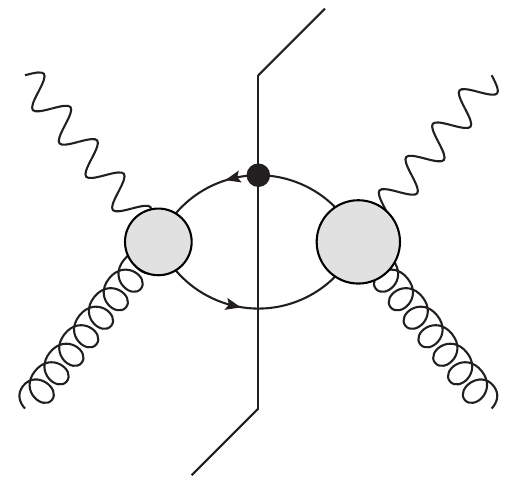}
\label{fig:FeynD}
}
\quad
\caption{Diagrams representing squared amplitudes for virtual Compton scattering from a parton to a two-parton final state, for a quark (or antiquark) initial state, with quark (or antiquark)~{\bf (a)} or gluon {\bf (b)} fragmentation (represented by the black circle), and for a gluon initial state with quark~{\bf (c)} or antiquark {\bf (d)} fragmentation. The grey blobs represent the possible connections within the diagrams, with the virtual photon ($q$), initial parton ($p$), fragmentating parton ($k_1$) and nonfragmenting parton ($k_2$) momenta labeled.}
\label{fig:Feyn.Diag}
\end{figure}

In calculating the partonic cross sections, it will be convenient to use the Mandelstam variables,
\begin{subequations}
\label{eq:mand}
\begin{eqnarray}
\label{eq:mand.s}
    s &=& (p+q)^2 = 2 p \cdot q - Q^2,
\\
\label{eq:mand.t}
    t &=& (q-k_1)^2 = -2 q \cdot k_1 - Q^2,
\\
\label{eq:mand.u}
    u &=& (p-k_1)^2 = -2 p \cdot k_1,
\end{eqnarray}
\end{subequations}
where we have taken the partons to be massless, with $q^2 = -Q^2$.
From Eqs.~(\ref{eq:mand}) it is straightforward to verify that 
    $s + t + u = -Q^2$.
When calculating the partonic cross sections, it is more natural to use partonic variables, which are more directly related to our parton momentum fractions and hadronic phase space variables.
In terms of these, the relevant Mandelstam variables are given by
\begin{subequations}
\begin{eqnarray}
\label{eq:s.to.part}
    s &=& \frac{(1-\hat{x})}{\hat{x}}\, Q^2, 
\\
\label{eq:t1.to.part}
    t &=& -Q^2 + \hat{z}\left(Q^2 - q_{T}^2 \right),
\\
\label{eq:u1.to.part}
    u &=& -\frac{\hat{z}}{\hat{x}}\, Q^2.
\end{eqnarray}
\end{subequations}
Also, when evaluating the spin-dependent cross sections, we use the spinor convention
\begin{align}
\label{eq:u-uB}
    u(p)\bar{u}(p) 
    &= \frac{1}{2} \big(1 + \widehat{\lambda}\gamma_5 \big)\slashed{p}
\end{align}
for quarks, and the polarization vector notation~\cite{deFlorian:1998fq},
\begin{align}
\label{eq:eps-epsB}
    \varepsilon^{\mu}(p)\varepsilon^{*\nu}(p) 
    &= \frac12\,
    \Big(\! -g^{\mu\nu}
    + \frac{i\widehat{\lambda}}{p \cdot X}\epsilon^{\mu\nu\alpha\beta}p_{\alpha}X_{\beta} \Big)
\end{align}
for gluons, where $\widehat{\lambda}$ is the helicity of the incoming parton (which is not necessarily the same as $\lambda$ defined above to be the initial state nucleon helicity), and $X$ is an arbitrary four vector which does not appear in the final squared amplitudes.

In our calculation of the unpolarized and polarized partonic differential cross sections we may bring the leptonic tensor inside the phase space integration of Eq.~(\ref{eq:part-tens.amps}) and contract directly with the scattering amplitudes calculated from the graphs in \fref{Feyn.Diag},
\begin{eqnarray}
\label{eq:lep-tens.contract.part-tens}
    L_{\mu\nu}\widehat{W}^{\mu\nu} = \int \diff \Pi \, {\cal H}_{ij},
\end{eqnarray}
where
    ${\cal H}_{ij} = L_{\mu\nu}{\cal M}^{\mu \dagger}_i{\cal M}^{\nu}_j$.
Note that we may write
\begin{eqnarray}
\label{eq:amplitude.separation}
    {\cal M}^{\mu \dagger}_i{\cal M}^\nu_j 
    = \big({\cal M}^{\mu \dagger}_i {\cal M}^{\nu}_j\big)^{\rm U} 
    + \widehat{\lambda}\,
      \big({\cal M}^{\mu \dagger}_i {\cal M}^{\nu}_j\big)^{\rm P},
\end{eqnarray}
where 
    $\rm U$
and
    $\rm P$
represent helicity-independent and helicity-dependent pieces of the scattering amplitude, respectively, since Eqs.~(\ref{eq:u-uB}) and (\ref{eq:eps-epsB}) contain terms independent of and linearly proportional to 
    $\widehat{\lambda}$.
It is then possible to write
\begin{eqnarray}
\label{eq:d}
    {\cal H}_{ij}
    = {\cal H}^{\rm U}_{ij} + \widehat{\lambda}\, {\cal H}^{\rm P}_{ij},
\end{eqnarray}
where
\begin{subequations}
\begin{eqnarray}
    {\cal H}^{\rm U}_{ij} 
    = L_{\mu\nu}\, \big( {\cal M}^{\mu \dagger}_i\, {\cal M}^\nu_j \big)^{\rm U},
\end{eqnarray}
and
\begin{eqnarray}
    {\cal H}^{\rm P}_{ij}
    = L_{\mu\nu}\, \big( {\cal M}^{\mu \dagger}_i\, {\cal M}^\nu_j \big)^{\rm P}.
\end{eqnarray}
\end{subequations}
The unpolarized contributions to 
    ${\cal H}_{ij}$
for each of the channels in Fig.~\ref{fig:Feyn.Diag} are given by
\begin{subequations}
\label{eq:upol}
\begin{eqnarray}
\label{eq:qq_channel_upol}
    \mathcal{H}^{\rm U}_{qq}
    &=& \frac{64 \pi \alpha_s e_q^2}{3 \hat{x} (1-\hat{x}) y^2 {\cal Q}_1^2}
    \Big[ 
        (1 + \hat{x}^2 \hat{z}^2) (1+\bar{y}^2)\, Q^4
        +\, 8 \hat{x}^2 \hat{z}^2 \bar{y}\, Q^2 q_T^2\,
\notag\\
    && \hspace*{3cm}
        +\, \hat{x}^2 \hat{z}^2 (1+\bar{y}^2)\, q_T^4
    \Big],
\\
    \mathcal{H}^{\rm U}_{qg}
    &=& \frac{64 \pi \alpha_s e_q^2}{3 (1-\hat{x}) y^2 {\cal Q}_2^2} 
    \Big[
        \Big( (2 + \hat{x}^2 \hat{z}^2) (1+\bar{y}^2)
                  - 4 \hat{x} \hat{z} \bar{y}
                  - 2 \hat{x} y^2 \big(1 - \hat{x} (1-\hat{z})\big)
        \Big) Q^4 
\notag\\
    && \hspace*{0.5cm}
        + 2 \hat{x} \hat{z} 
            \big( 4 \hat{x} \hat{z} \bar{y} + \hat{x} y^2 - 1 - \bar{y}^2
            \big)\, Q^2 q_T^2 
        + \hat{x}^2 \hat{z}^2 (1+\bar{y}^2)\, q_T^4\
    \Big],
\\
\label{eq:gq_channel_upol}
     \mathcal{H}^{\rm U}_{gq}
     &=& \frac{8 \pi \alpha_s e_q^2\, Q^2}{\hat{x} y^2 {\cal Q}_1^2 {\cal Q}_2^2} 
     \Big[ 
        \Big( (1 + 2 \hat{x}^2 \hat{z}^2) (1+\bar{y}^2)
                + 2 \hat{x}^2 y^2 (1-\hat{z})
                - 4 \hat{x} \hat{z} \bar{y}
                - 2 \hat{x} y^2 
        \Big)\, Q^4 
\notag\\
     && \hspace*{0.5cm}
        +\, 2 \hat{x} \hat{z}
            \big( 8 \hat{x} \hat{z} \bar{y}
                  + \hat{x} y^2
                  - 1 - \bar{y}^2
            \big)\, Q^2 q_T^2
        +\, 2 \hat{x}^2 \hat{z}^{2} (1+\bar{y}^2)\, q_T^4
     \Big],
\end{eqnarray}
\end{subequations}
for the $qq$, $qg$ and $gq$ channels, respectively, where $\alpha_s$ is the strong coupling constant.
For shorthand here we have defined
\begin{equation}
{\cal Q}_1^2 \equiv Q^2 (1 - \hat{z}) + \hat{z} q_T^2, \qquad
{\cal Q}_2^2 \equiv Q^2 \big(1-\hat{x} (1-\hat{z})\big)-\hat{x} \hat{z} q_T^2,
\end{equation}
and $\bar y \equiv 1-y$.
For the polarized case, the corresponding hard coefficients are given by 
\begin{subequations}
\label{eq:pol}
\begin{eqnarray}
\label{eq:qq_channel_pol}
     \mathcal{H}^{\rm P}_{qq}
     &=& -\frac{64 \pi \alpha_s e_q^2\, (2-y)}{3 \hat{x} (1-\hat{x}) y {\cal Q}_1^2} 
     \Big[
          (1 + \hat{x}^2 \hat{z}^2) Q^4 
        - \hat{x}^2 \hat{z}^2 q_T^4
     \Big],
\\
\label{eq:qg_channel_pol}
     \mathcal{H}^{\rm P}_{qg}
     &=& -\frac{64 \pi \alpha_s e_q^2\, \hat{x} (2-y)}{3 (1-\hat{x}) y {\cal Q}_2^2}
     \Big[
          \big( 2 + \hat{x} \hat{z}^{2} - 2 \hat{x} \hat{z} \big) Q^4 
        + 2 \hat{z} (1-\hat{x}) Q^2 q_T^2
        - \hat{x} \hat{z}^2 q_T^4
     \Big],~~~~~
\\
\label{eq:gq_channel_pol}
     \mathcal{H}^{\rm P}_{gq}
     &=&
     \frac{8 \pi \alpha_s e_q^2\, (2-y) Q^2}{\hat{x} y {\cal Q}_1^2 {\cal Q}_2^2}
     \Big[
        (2 \hat{x}^2 \hat{z}^2 - 2 \hat{x}^2 \hat{z} + 2 \hat{x} - 1)\, Q^4 
\notag\\
     && \hspace*{3cm}
        +\, 2 \hat{x} \hat{z} (1-\hat{x})\, Q^2 q_T^2
        - 2 \hat{x}^2 \hat{z}^2\, q_T^4
     \Big].
\end{eqnarray}
\end{subequations}
These results are consistent with those obtained previously by Koike {\it et al.}~\cite{Koike:2006fn}.
Furthermore, since we neglect parton masses in our calculations, the expressions for both the 
    $\gamma g \to q\bar{q}$
and 
    $\gamma g \to \bar{q}q$
channels of Fig.~\ref{fig:Feyn.Diag}(c)--\ref{fig:Feyn.Diag}(d) are given by Eqs.~(\ref{eq:gq_channel_upol}) and (\ref{eq:gq_channel_pol}).
With these results for the hard coefficients, we can proceed to compute the physical cross sections for the polarized SIDIS production of a hadron $h$, which we turn to in the next section.

\section{Numerical simulations}
\label{sec:results}

Typically in experiments involving polarization of initial or final state particles, one defines asymmetries between cross sections for spin-aligned and anti-aligned configurations.
The relevant asymmetry for the polarized semi-inclusive electroproduction of a hadron $h$ can be defined in terms of the differential cross sections $\diff \sigh$ and $\diff \Delta \sigh$ from Sec.~\ref{ssec:definitions},
\begin{eqnarray}
\label{eq:asymmetry}
    A_{LL}^h = \frac{\diff \Delta \sigh}{\diff \sigh}.
\end{eqnarray}
The results presented in this section will focus on the SIDIS process for a polarized proton with production of charged pions in the final state at energies accessible at the current 12~GeV Jefferson Lab, a possible future 20--24~GeV upgrade, and EIC kinematics.

\subsection{Phase space restrictions}
\label{ss:phase-space-restrictions}

The kinematic phase space for the SIDIS reaction is four dimensional.
We must impose boundaries for each of the kinematic variables $x$, $y$, $z$ and $\PhT$ to take into account theoretical and physical conditions as well as constraints imposed by relevant experimental conditions.
For the fragmentation variable, following previous analyses of SIDIS data \cite{Sato:2016wqj, Moffat:2021dji}, we impose the restriction
    $0.2 < z < 0.8$,
over which good descriptions of data in terms of leading twist fragmentation functions can be obtained within the JAM global QCD analysis framework.

For the four-momentum transfer squared, 
    $Q^2$,
we impose a lower bound at the mass of the charm quark,
    $Q^2 > m_c^2 = (1.28~{\rm GeV})^2$.
Furthermore, imposing a cut on the invariant mass squared of the final hadronic state,
    $W^2 = (P+q)^2 = M^2 + Q^2 (1-x)/x > W_{\rm min}^2 = 4~{\rm GeV}^2$,
we obtain the further restriction,
    $Q^2 > (W_{\rm min}^2-M^2) x/(1-x)$.
For EIC kinematics with center of mass energy $\sqrt{s} = 140$~GeV, at each value of $x$ the minimum allowed value of $Q^2$ is the maximum of these two conditions.
For Jefferson Lab kinematics at a current center of mass energy $\sqrt{s} = 4.8$~GeV and a possible future energy of $\sqrt{s} = 6.5$~GeV (corresponding to laboratory frame electron beam energies of 12 and 22~GeV, respectively), we impose a third condition on $Q^2$, noting that the CLAS12 detector can detect electrons scattered at angles as low as  $\theta \gtrsim \theta_{\rm min} \approx 5^\circ$,
which implies
\begin{eqnarray}
\label{eq:Q2-min-angle}
    Q^2 > \frac{2 E^2 (1-\cos{\theta_{\rm min}})}{1 + (E/M x)(1 - \cos{\theta_{\rm min}})},
\end{eqnarray}
where
    $\theta_{\rm min}$
is the minimum scattering angle and
    $E$
is the energy of the electron beam in the laboratory frame.
At these kinematics the minimum value of $Q^2$ is then the maximum of these three values at each $x$.
A strict upper bound for $Q^2$ at each $x$ is obtained simply by requiring that $y < 1$, giving
\begin{eqnarray}
    Q^2 < Q^2_{\rm max} = (s - M^2) x.
\end{eqnarray}
The values of 
    $x_{\rm min}$
and 
    $x_{\rm max}$
are defined as the points where the $Q^2$ range shrinks to zero,
    $Q^2_{\rm min} = Q^2_{\rm max}$.
In the neighborhood of 
    $x_{\rm min}$,
we have
    $Q_{\rm min}^2 = m_c^2$,
so that
\begin{eqnarray}
x_{\rm min} = \frac{m_c^2}{(s - M^2)},
\end{eqnarray}
while in the neighborhood of
    $x_{\rm max}$,
we have
    $Q_{\rm min}^2 = (W^2-M^2)\, x/(1-x)$,
so that
\begin{eqnarray}
x_{\rm max} = \frac{s - W^2}{s - M^2}.
\end{eqnarray}

Finally, to avoid regions of low hadron transverse momenta, where a TMD description would be more appropriate, we require
    $q_T \gtrsim Q$.
An upper bound on 
    $q_T$
is derived by requiring the mass of the unobserved hadronic final state, 
    $W_{\mbox{\tiny \rm SIDIS}}^2 = (P + q - \Ph)^2$,
to be at least the proton mass.
This gives
\begin{equation}
\label{eq:W2-SIDIS-qT}
    W_{\mbox{\tiny \rm SIDIS}}^2 
    = M^2 + \Mh^2 + \frac{Q^2 \left(1 - x - z\right)}{x}
     + \frac{2 Q^2 z}{\rho^2-1} 
        \Bigg( \rho\ \sqrt{1 - (\rho^2-1) \frac{\MhT^2}{Q^2 z^2}} - 1
        \Bigg),
\end{equation}
where $\MhT^2 \equiv M^2 + \PhT^2$ is the transverse mass squared of the produced hadron, and we define $\rho^2 \equiv 1 + 4 M^2 x^2/Q^2$.
Solving the inequality 
    $W_{\rm SIDIS}^2 > M^2$
for $q_T/Q$ gives then a condition on the maximal value of $q_T$ relative to $Q$,
\begin{eqnarray}
\frac{q_T}{Q}\
<\
\sqrt{
\frac{1}{\rho^2-1}
  \bigg[ 1 - (\rho^2-1) \frac{M^2}{Q^2 z^2} \bigg]
- \frac{(\rho^2-1)}{4 \rho^2 z^2}
  \bigg[ \frac{1-x-z}{x} + \frac{\Mh^2}{Q^2} - \frac{2z}{\rho^2-1} \bigg]^2
}.
\end{eqnarray}

\subsection{Statistical errors and kinematic bins}
\label{ss:statistical-errors-and-kinematic-bins}

The statistical uncertainty for the double-spin asymmetry, defined in Eq.~(\ref{eq:asymmetry}), is given by
\begin{eqnarray}
\label{eq:A_LL-stat-error}
    \delta A^{h}_{LL} = \sqrt{\frac{1 + (A^{h}_{LL})^2}{N}}
    \approx \frac{1}{\sqrt{N}} \quad {\rm if}\ A^{h}_{LL} \ll 1,
\end{eqnarray}
where 
    $N$
is the number of events in the kinematic bin in which the asymmetry is calculated, and is related to the unpolarized cross section $\sigh$ by
\begin{eqnarray}
\label{eq:N-to-xsec}
\begin{aligned}
    N = \mathcal{L}\, \sigh
     &= \mathcal{L} \int_{\rm bin} \diff x\, \diff y\, \diff z\, \diff \PhT^2\, \frac{\diff \sigh}{\diff x\, \diff y\, \diff z\, \diff \PhT^2} \\
    &\approx \mathcal{L}\, \Delta x\, \Delta y\, \Delta z\, \Delta \PhT^2
    \bigg( \frac{\diff \sigh}{\diff x\, \diff y\, \diff z\, \diff \PhT^2} \bigg)_{\rm center},
\end{aligned}
\end{eqnarray}
where 
    $\mathcal{L}$
is the integrated luminosity for the scattering events.
The approximation made for the integral giving the total unpolarized cross section is valid assuming that the differential unpolarized cross section is roughly constant in a given bin.

The kinematic bins are then constructed by first generating a $4 \times 4$ grid as
    $[x_{\rm min},x_{\rm max}] \times [Q_{\rm min}^2,Q_{\rm max}^2] \times [z_{\rm min},z_{\rm max}] \times [q_{T,\rm min},q_{T,\rm max}]$
for the absolute minima and maxima for each kinematic variable at a given $\sqrt{s}$ value.
The $z$ domain is divided into 10 bins, and the $q_T$ range is divided into bins of width $1~{\rm GeV}$.
The $x$ and $Q^2$ ranges are divided into an equal number of bins, $n_{\rm bins}$, with equal widths in log-space,
\begin{eqnarray}
     n_{\rm bins} = \frac{1}{\Delta (\log{x})}\log\Big( \frac{x_{\rm max}}{x_{\rm min}} \Big)
,\end{eqnarray}
rounded up to the next integer.
For the bin width we choose $\Delta (\log{x}) = 0.1$.
The domain is then truncated such that the midpoint of each bin is kinematically valid with respect to the phase space restrictions outlined in~\ssref{phase-space-restrictions}.

\begin{figure}
\includegraphics[width=\textwidth]{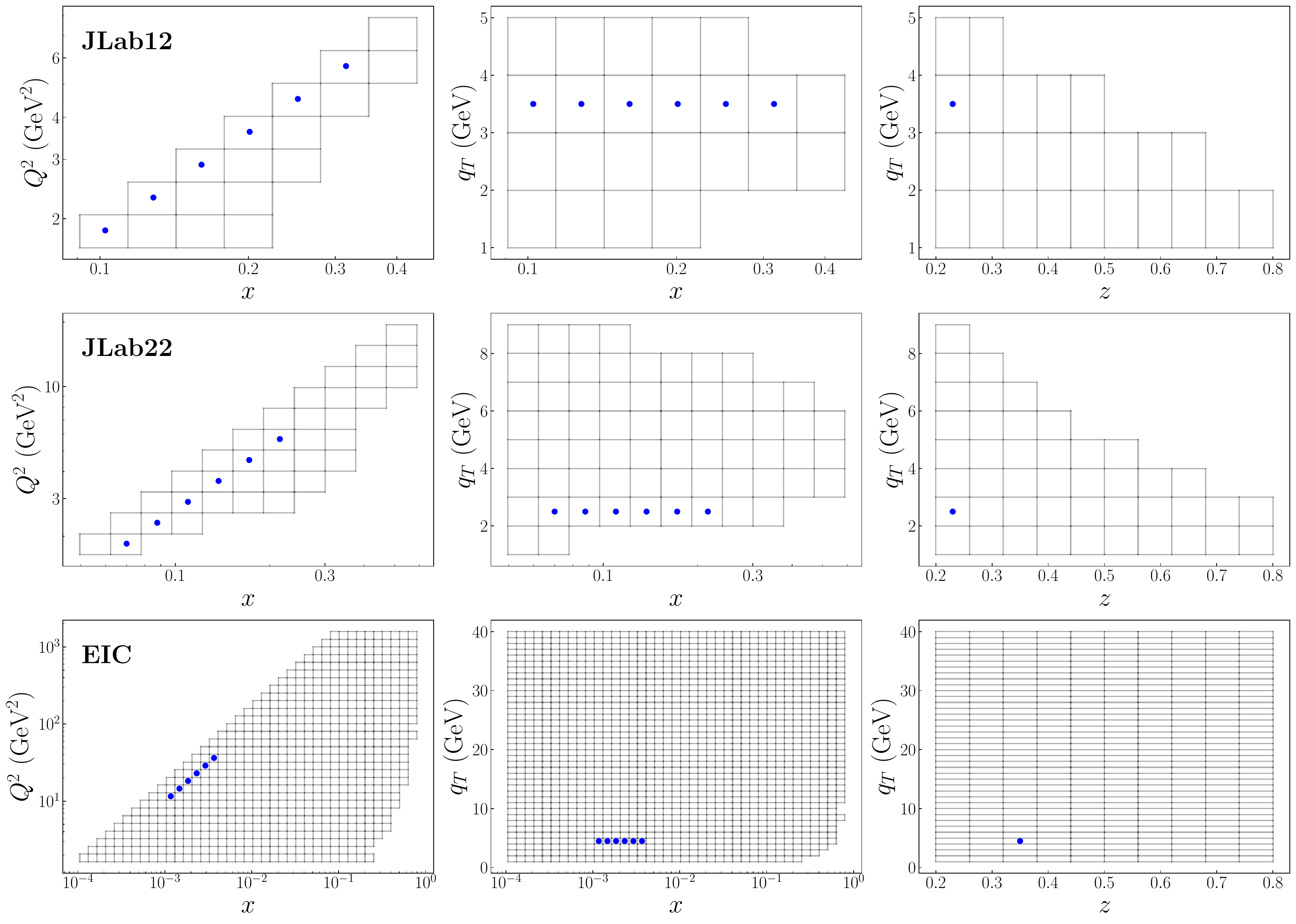}
\caption{Available kinematics at Jefferson Lab with a 12~GeV (top row) and 22~GeV (middle row) electron beam and at the EIC (bottom row) with $\sqrt{s}=140$~GeV.  The boxes indicate the available kinematic bins, as discussed in \ssref{statistical-errors-and-kinematic-bins}, and the blue dots represent the centers of the bins at which the asymmetries are calculated in our analysis (see \fref{ALL-plots}).
}
\label{fig:ALL-kinematics}
\end{figure}

Plots of these allowed bins in the $(x,Q^2)$, $(x,q_T)$, and $(z,q_T)$ planes relevant for present and future facilities are shown in \fref{ALL-kinematics}.
Displayed are the kinematics for Jefferson Lab with the current 12~GeV electron beam ($\sqrt{s} = 4.8$~GeV), and for a possible future 22~GeV electron beam energy ($\sqrt{s} = 6.5$~GeV), as well as for the planned EIC ($\sqrt{s} = 140$~GeV). 
The asymmetry $A_{LL}^{\pi^+}$ is calculated at the center of the kinematic bins, indicated in \fref{ALL-kinematics} by the blue dots at the centers of each of the bins (corresponding approximately to best case scenarios), along with statistical uncertainties for the asymmetry given by \eref{A_LL-stat-error}.
Essentially, the size of the asymmetry and uncertainties associated with the asymmetry are simultaneously maximized and minimized, respectively, in a kinematic neighborhood of the bins for each chosen center of mass energy in \fref{ALL-kinematics}.

\begin{figure}
\includegraphics[height=0.75\textheight]{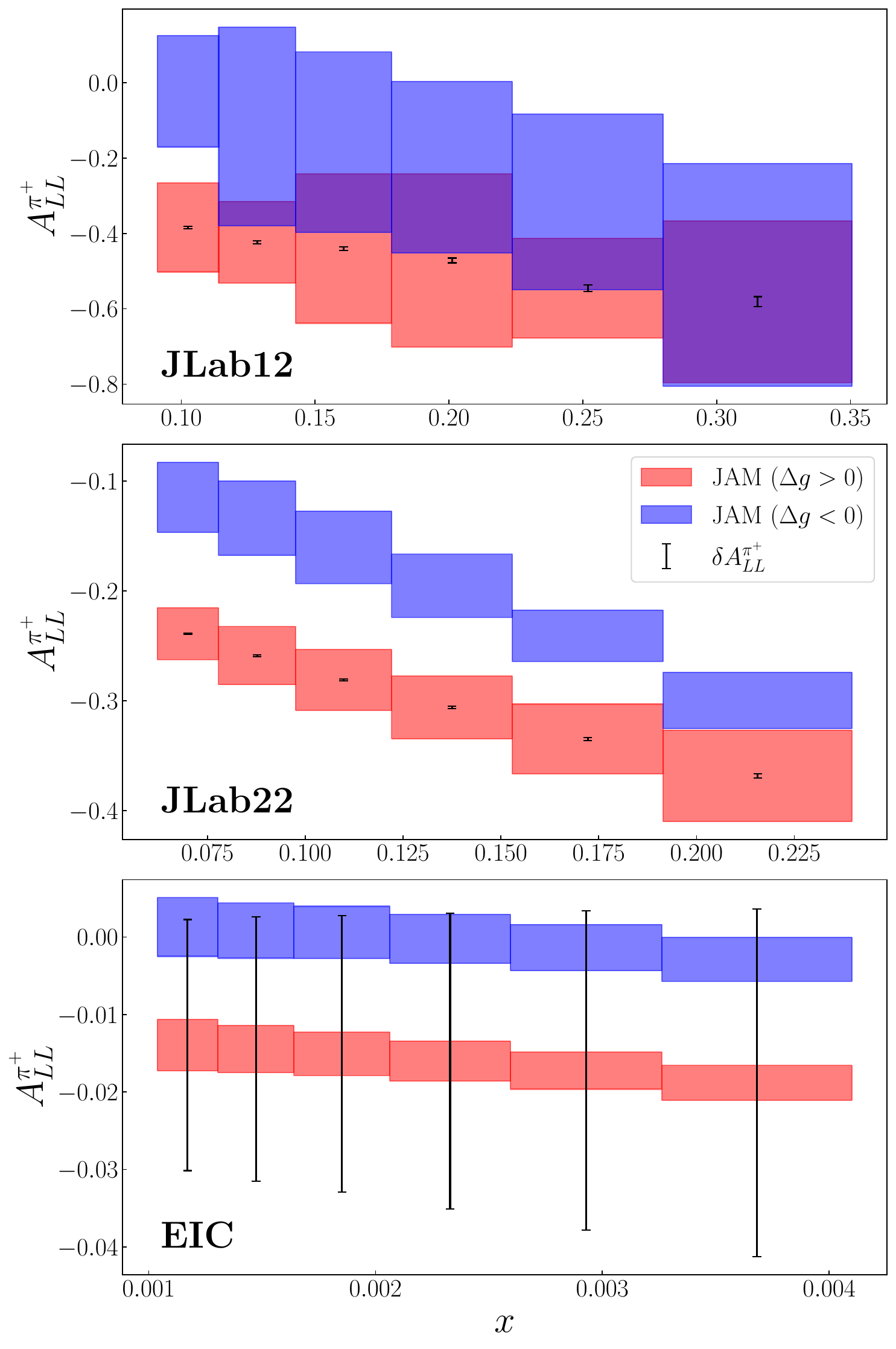}
\caption{Double longitudinal spin asymmetry $A_{LL}^{\pi^+}$ for semi-inclusive $\pi^+$ production from a proton, at kinematics indicated in \fref{ALL-kinematics} at Jefferson Lab (JLab) with both a 12~GeV and 22~GeV electron beam and at the EIC with $\sqrt{s}=140~{\rm GeV}$. Note that the heights of the colored boxes give a $1\sigma$ uncertainty in the asymmetry from the PDF replicas, while the error bars give the statistical uncertainty of the asymmetry from Eq.~(\ref{eq:A_LL-stat-error}).
}
\label{fig:ALL-plots}
\end{figure}

In Fig.~\ref{fig:ALL-plots} we show the projected $A_{LL}^{\pi^+}$ asymmetries at the Jefferson Lab 12~GeV and 22~GeV, and EIC, kinematics.
For the statistical uncertainties on the Jefferson Lab projections, we take a luminosity of $\diff \mathcal{L} / \diff t = 10^{-35}~{\rm cm}^{-2}{\rm s}^{-1}$,
which for 10 days of running would correspond to an integrated luminosity of $\approx 86$~fb$^{-1}$.
For the EIC statistical uncertainties, we assume an integrated luminosity of $\mathcal{L} = 10~{\rm fb}^{-1}$, which is expected to be the achievable luminosity for SIDIS experiments such as those presented here~\cite{AbdulKhalek:2021gbh}.
The asymmetries produced at Jefferson Lab 12~GeV kinematics are relatively large, while their statistical uncertainties are quite small compared to the scale of the asymmetries.
For most kinematics, the asymmetry bands with positive and negative polarized gluons overlap significantly, which would make it difficult to discriminate between the positive and negative $\Delta g$ solutions.
The separation between the positive and negative bands becomes clearer at smaller $x$ values. However, the current 12~GeV electron energy restricts the range of $x$ down to which the asymmetries can be probed.

A larger portion of the intermediate- and low-$x$ region can be accessed, on the other hand, with an energy upgraded 22~GeV electron beam.
At this energy, and with the same luminosity, the asymmetry is still sizable and retains the small statistical uncertainties as for the 12~GeV case.
Therefore, significantly better discrimination between the positive and negative asymmetry bands is found in this case, with a clearer separation between the means and the boundaries of the $1\sigma$ region of the two bands.

At the EIC, on the other hand, the calculated double spin asymmetries are some one to two orders of magnitude smaller than the largest asymmetries at Jefferson Lab, for either the 12~GeV  or 22~GeV beam energies.
Furthermore, with the lower available luminosity, the statistical uncertainties on the asymmetry are quite large relative to the size of the asymmetries and the separation between the positive and negative $\Delta g$ bands.

These two observed behaviors are intimately related to each other through the scaling of the asymmetry with the center of mass energy, $\sqrt{s}$.
In the convolution integrals of Eqs.~(\ref{eq:fact.thm}), the PDFs and fragmentation functions do not depend on $\sqrt{s}$, so the scale dependence of the unpolarized and polarized cross sections can only arise from the hard scattering amplitudes in Eqs.~(\ref{eq:upol}) and (\ref{eq:pol}) or the kinematic factors multiplying the product between the leptonic and hadronic tensors in \eref{upol.xsec}.
In fact, both of these depend on the lepton inelasticity $y$, which is related to the center of mass energy by
    $y = Q^2/[x(s-M^2)]$.
Since the kinematic factors in \eref{upol.xsec} are the same for both the unpolarized and polarized cross sections, they naturally cancel in the asymmetry.
However, it is clear from the unpolarized and polarized hard scattering amplitudes  that the asymmetry possesses a strong scale dependence.
In the unpolarized expressions there is a common factor of $1/y^2$, while in the polarized expressions the common factor is $(2-y)/y$, giving an overall factor $(2-y)y$ for the asymmetry.

\begin{figure}
\includegraphics[width=0.7\textwidth]{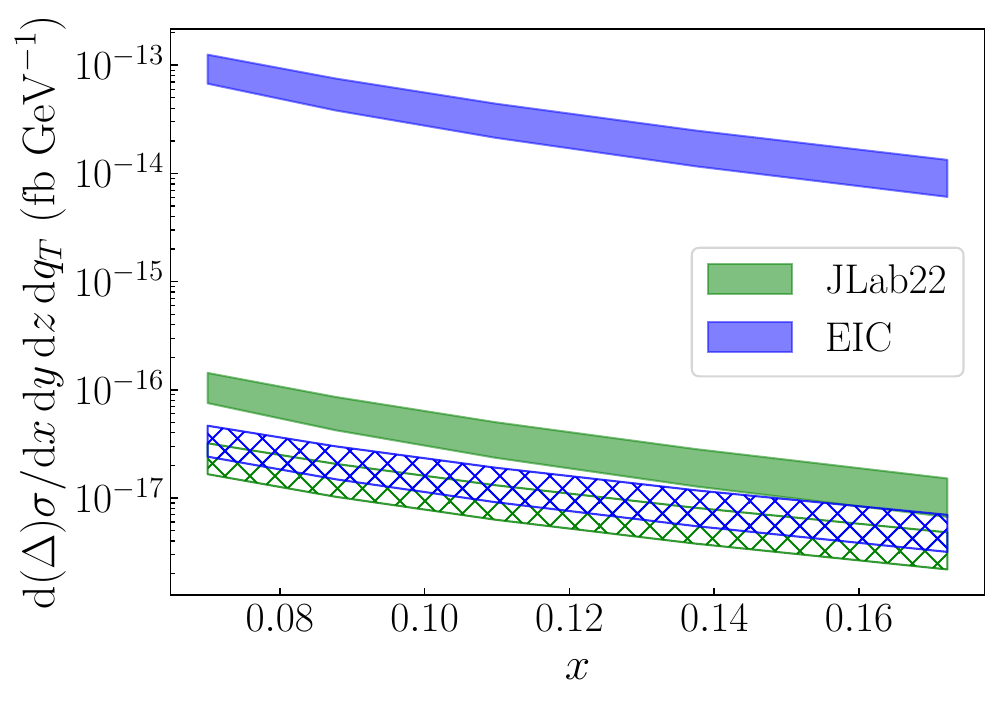}
\caption{Unpolarized (solid bands) and polarized (hatched bands) differential cross sections calculated for semi-inclusive $\pi^+$ production at the $(x,Q^2,z,q_T)$ kinematics displayed in \fref{ALL-kinematics} for Jefferson Lab with a 22~GeV beam energy (green bands) and EIC center of mass energy $\sqrt{s}=140$~GeV (blue bands).}
\label{fig:cross-section-scale}
\end{figure}

A comparison of the unpolarized and polarized SIDIS $\pi^+$ production cross sections, calculated at the Jefferson Lab 22~GeV and EIC kinematics, is shown in \fref{cross-section-scale}.
The polarized cross section here displays a relatively weak dependence on $\sqrt{s}$, while the unpolarized cross section depends strongly on $\sqrt{s}$, increasing with larger center of mass energies.
Combined, these behaviors act to suppresses the double spin asymmetry at larger $\sqrt{s}$.
The size of the statistical uncertainties for the EIC asymmetry in \fref{ALL-plots} can also be understood by considering the relative uncertainty of the asymmetry, which is given as
\begin{eqnarray}
    \label{eq:ALL-rel-unc}
    \frac{\delta A_{LL}^{h}}{A_{LL}^h}\,
    =\, \frac{1}{A_{LL}^{h}\sqrt{\mathcal{L} \sigma}}\,
    =\, \frac{1}{\Delta \sigma} \sqrt{\frac{\sigma}{\mathcal{L}}}\,
.\end{eqnarray}
It immediately follows then that the scaling of the relative uncertainty of the asymmetry is dominated by the scaling of $\sqrt{\sigma}$ with $\sqrt{s}$.

While a $1\%-5\%$ asymmetry is certainly measurable, at luminosities of the order of 10~fb$^{-1}$ the resolution of the positive and negative asymmetry bands becomes quite poor.
In order to discriminate between the two sets of gluon PDFs at low $x$, one would require integrated luminosities of at least one order of magnitude larger than what is currently projected for the SIDIS process at the EIC.

\section{Conclusion}
\label{sec:conc}

The proton spin decomposition has remained a fundamental challenge for nuclear physics for over 3 decades, even with the remarkable progress made on both the experimental and theoretical sides of the problem.
The quest for its understanding at the QCD level continues to generate surprises, such as the recent realization that the conventional picture of a positive gluon polarization may hinge on theoretical assumptions about PDF positivity~\cite{Zhou:2022wzm}, the relaxation of which reveals the possibility of solutions with negative $\Delta g$ that describe existing data equally well.

In this study we have considered the double spin asymmetry for charged pions produced with large transverse momentum in the polarized SIDIS process, which is particularly sensitive to the gluon channel, as a means of discriminating between positive and negative $\Delta g$ distributions.
In particular, we compared the constraining power of current and future electron scattering facilities, including Jefferson Lab and the EIC, and found that an $\approx 20$~GeV beam at the high luminosity Jefferson Lab is especially well-suited for the discrimination between the positive and negative solutions.
Because of the scaling behavior of the asymmetry with $\sqrt{s}$, significantly larger integrated luminosities would be required at EIC energies to overcome the suppression of the asymmetry and relatively large statistical uncertainties at these higher energies.
Further work is needed to understand which observables will allow maximal utilization of the EIC's reach into low-$x$ kinematics, and ability to constrain $\Delta g$ in this region through future global QCD analyses~\cite{AbdulKhalek:2021gbh, Zhou:2021llj}.

Independent pathways towards constraining the polarization of the glue could involve inclusive charm meson production in polarized DIS through the photon-gluon fusion process~\cite{COMPASS:2012mpe, Anderle:2021hpa}.
Alternatively, lattice data on matrix elements that are sensitive to the shape and sign of $\Delta g$  \cite{HadStruc:2022yaw} may be fitted simultaneously with experimental data in future global QCD analyses in order to obtain a consistent picture of the role of gluon polarization in the proton spin decomposition.


\begin{acknowledgements}

This work was supported by the U.S. Department of Energy Contract No.~DE-AC05-06OR23177, under which Jefferson Science Associates, LLC operates Jefferson Lab.
R.M.W. was partially supported by the Jefferson Science Associates Initiatives Fund for Undergraduate Student Support; by the U.S. Department of Energy, Office of Science, Office of Workforce Development for Teachers and Scientists (WDTS) under the Science Undergraduate Laboratory Internships Program (SULI); and by the National Science Foundation under Grant No.~1950141.
The work of N.S. was supported by the DOE, Office of Science, Office of Nuclear Physics in the Early Career Program.
W.M. was supported in part by the Southgate Fellowship, and by the Australian Research Council through the Centre of Excellence for Dark Matter Particle Physics (CE200100008).
The work of Y. Z. is supported by the Guangdong Major Project of Basic and Applied Basic Research No. 2020B0301030008, the National Natural Science Foundation of China under Grants No. 12022512, No. 12035007.

\end{acknowledgements}

\bibliography{bibliography}

@article{Aidala:2012mv,
    author = "Aidala, Christine A. and Bass, Steven D. and Hasch, Delia and Mallot, Gerhard K.",
    title = "{The Spin Structure of the Nucleon}",
    eprint = "1209.2803",
    archivePrefix = "arXiv",
    primaryClass = "hep-ph",
    doi = "10.1103/RevModPhys.85.655",
    journal = "Rev. Mod. Phys.",
    volume = "85",
    pages = "655--691",
    year = "2013"
}

@article{STAR:2014wox,
    author = "Adamczyk, L. and others",
    collaboration = "STAR",
    title = "{Precision Measurement of the Longitudinal Double-spin Asymmetry for Inclusive Jet Production in Polarized Proton Collisions at $\sqrt{s}=200$ GeV}",
    eprint = "1405.5134",
    archivePrefix = "arXiv",
    primaryClass = "hep-ex",
    doi = "10.1103/PhysRevLett.115.092002",
    journal = "Phys. Rev. Lett.",
    volume = "115",
    number = "9",
    pages = "092002",
    year = "2015"
}

@article{deFlorian:2014yva,
    author = "de Florian, Daniel and Sassot, Rodolfo and Stratmann, Marco and Vogelsang, Werner",
    title = "{Evidence for polarization of gluons in the proton}",
    eprint = "1404.4293",
    archivePrefix = "arXiv",
    primaryClass = "hep-ph",
    doi = "10.1103/PhysRevLett.113.012001",
    journal = "Phys. Rev. Lett.",
    volume = "113",
    number = "1",
    pages = "012001",
    year = "2014"
}

@article{STAR:2019yqm,
    author = "Adam, J. and others",
    collaboration = "STAR",
    title = "{Longitudinal double-spin asymmetry for inclusive jet and dijet production in pp collisions at $\sqrt{s} = 510$ GeV}",
    eprint = "1906.02740",
    archivePrefix = "arXiv",
    primaryClass = "hep-ex",
    doi = "10.1103/PhysRevD.100.052005",
    journal = "Phys. Rev. D",
    volume = "100",
    number = "5",
    pages = "052005",
    year = "2019"
}

@article{STAR:2021mfd,
    author = "Abdallah, M. S. and others",
    collaboration = "STAR",
    title = "{Longitudinal double-spin asymmetry for inclusive jet and dijet production in polarized proton collisions at $\sqrt{s}=200$ GeV}",
    eprint = "2103.05571",
    archivePrefix = "arXiv",
    primaryClass = "hep-ex",
    doi = "10.1103/PhysRevD.103.L091103",
    journal = "Phys. Rev. D",
    volume = "103",
    number = "9",
    pages = "L091103",
    year = "2021"
}

@article{STAR:2021mqa,
    author = "Abdallah, M. S. and others",
    collaboration = "STAR",
    title = "{Longitudinal double-spin asymmetry for inclusive jet and dijet production in polarized proton collisions at $\sqrt{s}=510$ GeV}",
    eprint = "2110.11020",
    archivePrefix = "arXiv",
    primaryClass = "hep-ex",
    doi = "10.1103/PhysRevD.105.092011",
    journal = "Phys. Rev. D",
    volume = "105",
    number = "9",
    pages = "092011",
    year = "2022"
}

@article{PHENIX:2010aru,
    author = "Adare, A. and others",
    collaboration = "PHENIX",
    title = "{Event Structure and Double Helicity Asymmetry in Jet Production from Polarized $p+p$ Collisions at $\sqrt{s} = 200${\textasciitilde}GeV}",
    eprint = "1009.4921",
    archivePrefix = "arXiv",
    primaryClass = "hep-ex",
    doi = "10.1103/PhysRevD.84.012006",
    journal = "Phys. Rev. D",
    volume = "84",
    pages = "012006",
    year = "2011"
}

@article{Zhou:2022wzm,
    author = "Zhou, Y. and Sato, N. and Melnitchouk, W.",
    title = "{How well do we know the gluon polarization in the proton?}",
    eprint = "2201.02075",
    archivePrefix = "arXiv",
    primaryClass = "hep-ph",
    reportNumber = "JLAB-THY-22-3462",
    doi = "10.1103/PhysRevD.105.074022",
    journal = "Phys. Rev. D",
    volume = "105",
    number = "7",
    pages = "074022",
    year = "2022"
}

@article{Bass:2009ed,
    author = "Bass, Steven D. and Thomas, Anthony W.",
    title = "{The nucleon's octet axial-charge $g_A^{(8)}$ with chiral corrections}",
    eprint = "0912.1765",
    archivePrefix = "arXiv",
    primaryClass = "hep-ph",
    doi = "10.1016/j.physletb.2010.01.008",
    journal = "Phys. Lett. B",
    volume = "684",
    pages = "216--220",
    year = "2010"
}

@article{Ethier:2017zbq,
    author = "Ethier, J. J. and Sato, N. and Melnitchouk, W.",
    title = "{First simultaneous extraction of spin-dependent parton distributions and fragmentation functions from a global QCD analysis}",
    eprint = "1705.05889",
    archivePrefix = "arXiv",
    primaryClass = "hep-ph",
    reportNumber = "JLAB-THY-17-2464",
    doi = "10.1103/PhysRevLett.119.132001",
    journal = "Phys. Rev. Lett.",
    volume = "119",
    number = "13",
    pages = "132001",
    year = "2017"
}

@article{Candido:2020yat,
    author = "Candido, Alessandro and Forte, Stefano and Hekhorn, Felix",
    title = "{Can $ \overline{\mathrm{MS}} $ parton distributions be negative?}",
    eprint = "2006.07377",
    archivePrefix = "arXiv",
    primaryClass = "hep-ph",
    reportNumber = "TIF-UNIMI-2020-9",
    doi = "10.1007/JHEP11(2020)129",
    journal = "JHEP",
    volume = "11",
    pages = "129",
    year = "2020"
}

@article{Collins:2021vke,
    author = "Collins, John and Rogers, Ted C. and Sato, Nobuo",
    title = "{Positivity and renormalization of parton densities}",
    eprint = "2111.01170",
    archivePrefix = "arXiv",
    primaryClass = "hep-ph",
    reportNumber = "JLAB-THY-21-3507",
    doi = "10.1103/PhysRevD.105.076010",
    journal = "Phys. Rev. D",
    volume = "105",
    number = "7",
    pages = "076010",
    year = "2022"
}

@article{Jager:2003ch,
    author = "Jager, Barbara and Stratmann, Marco and Kretzer, Stefan and Vogelsang, Werner",
    title = "{QCD hard scattering and the sign of the spin asymmetry A**pi(LL)}",
    eprint = "hep-ph/0310197",
    archivePrefix = "arXiv",
    reportNumber = "BNL-NT-03-32, RBRC-359",
    doi = "10.1103/PhysRevLett.92.121803",
    journal = "Phys. Rev. Lett.",
    volume = "92",
    pages = "121803",
    year = "2004"
}

@article{PHENIX:2015fxo,
    author = "Adare, A. and others",
    collaboration = "PHENIX",
    title = "{Inclusive cross section and double-helicity asymmetry for $\pi^{0}$ production at midrapidity in $p+p$ collisions at $\sqrt{s}=510$ GeV}",
    eprint = "1510.02317",
    archivePrefix = "arXiv",
    primaryClass = "hep-ex",
    doi = "10.1103/PhysRevD.93.011501",
    journal = "Phys. Rev. D",
    volume = "93",
    number = "1",
    pages = "011501",
    year = "2016"
}

@article{PHENIX:2014axc,
    author = "Adare, A. and others",
    collaboration = "PHENIX",
    title = "{Charged-pion cross sections and double-helicity asymmetries in polarized $p+p$ collisions at $\sqrt{s}$=200 GeV}",
    eprint = "1409.1907",
    archivePrefix = "arXiv",
    primaryClass = "hep-ex",
    doi = "10.1103/PhysRevD.91.032001",
    journal = "Phys. Rev. D",
    volume = "91",
    number = "3",
    pages = "032001",
    year = "2015"
}

@article{PHENIX:2020trf,
    author = "Acharya, U. A. and others",
    collaboration = "PHENIX",
    title = "{Measurement of charged pion double spin asymmetries at midrapidity in longitudinally polarized $p+p$ collisions at $\sqrt {s}$ = 510 GeV}",
    eprint = "2004.02681",
    archivePrefix = "arXiv",
    primaryClass = "hep-ex",
    doi = "10.1103/PhysRevD.102.032001",
    journal = "Phys. Rev. D",
    volume = "102",
    number = "3",
    pages = "032001",
    year = "2020"
}

@article{Wang:2019bvb,
    author = "Wang, B. and Gonzalez-Hernandez, J. O. and Rogers, T. C. and Sato, N.",
    title = "{Large Transverse Momentum in Semi-Inclusive Deeply Inelastic Scattering Beyond Lowest Order}",
    eprint = "1903.01529",
    archivePrefix = "arXiv",
    primaryClass = "hep-ph",
    reportNumber = "JLAB-THY-19-2897",
    doi = "10.1103/PhysRevD.99.094029",
    journal = "Phys. Rev. D",
    volume = "99",
    number = "9",
    pages = "094029",
    year = "2019"
}

@article{deFlorian:1998fq,
    author = "de Florian, D. and Vogelsang, W.",
    title = "{Next-to-leading order QCD corrections to inclusive hadron photoproduction in polarized lepton proton collisions}",
    eprint = "hep-ph/9712273",
    archivePrefix = "arXiv",
    reportNumber = "CERN-TH-97-280",
    doi = "10.1103/PhysRevD.57.4376",
    journal = "Phys. Rev. D",
    volume = "57",
    pages = "4376--4384",
    year = "1998"
}

@article{Koike:2006fn,
    author = "Koike, Yuji and Nagashima, Junji and Vogelsang, Werner",
    title = "{Resummation for polarized semi-inclusive deep-inelastic scattering at small transverse momentum}",
    eprint = "hep-ph/0602188",
    archivePrefix = "arXiv",
    reportNumber = "BNL-NT-06-9, RBRC-586",
    doi = "10.1016/j.nuclphysb.2006.03.009",
    journal = "Nucl. Phys. B",
    volume = "744",
    pages = "59--79",
    year = "2006"
}

@article{Sato:2016wqj,
    author = "Sato, Nobuo and Ethier, J. J. and Melnitchouk, W. and Hirai, M. and Kumano, S. and Accardi, A.",
    title = "{First Monte Carlo analysis of fragmentation functions from single-inclusive $e^+ e^-$ annihilation}",
    eprint = "1609.00899",
    archivePrefix = "arXiv",
    primaryClass = "hep-ph",
    reportNumber = "JLAB-THY-16-2327, KEK-TH-1920, J-PARC-TH-0060",
    doi = "10.1103/PhysRevD.94.114004",
    journal = "Phys. Rev. D",
    volume = "94",
    number = "11",
    pages = "114004",
    year = "2016"
}

@article{Moffat:2021dji,
    author = "Moffat, Eric and Melnitchouk, Wally and Rogers, T. C. and Sato, Nobuo",
    title = "{Simultaneous Monte~Carlo analysis of parton densities and fragmentation functions}",
    eprint = "2101.04664",
    archivePrefix = "arXiv",
    primaryClass = "hep-ph",
    reportNumber = "JLAB-THY-21-3304",
    doi = "10.1103/PhysRevD.104.016015",
    journal = "Phys. Rev. D",
    volume = "104",
    number = "1",
    pages = "016015",
    year = "2021"
}

@article{AbdulKhalek:2021gbh,
    author = "Abdul Khalek, R. and others",
    title = "{Science Requirements and Detector Concepts for the Electron-Ion Collider}: {EIC Yellow Report}",
    eprint = "2103.05419",
    archivePrefix = "arXiv",
    primaryClass = "physics.ins-det",
    reportNumber = "BNL-220990-2021-FORE, JLAB-PHY-21-3198, LA-UR-21-20953",
    doi = "10.1016/j.nuclphysa.2022.122447",
    journal = "Nucl. Phys. A",
    volume = "1026",
    pages = "122447",
    year = "2022"
}

@article{Zhou:2021llj,
    author = "Zhou, Y. and Cocuzza, C. and Delcarro, F. and Melnitchouk, W. and Metz, A. and Sato, N.",
    title = "{Revisiting quark and gluon polarization in the proton at the EIC}",
    eprint = "2105.04434",
    archivePrefix = "arXiv",
    primaryClass = "hep-ph",
    reportNumber = "JLAB-THY-21-3368",
    doi = "10.1103/PhysRevD.104.034028",
    journal = "Phys. Rev. D",
    volume = "104",
    number = "3",
    pages = "034028",
    year = "2021"
}

@article{COMPASS:2012mpe,
    author = "Adolph, C. and others",
    collaboration = "COMPASS",
    title = "{Leading and Next-to-Leading Order Gluon Polarization in the Nucleon and Longitudinal Double Spin Asymmetries from Open Charm Muoproduction}",
    eprint = "1211.6849",
    archivePrefix = "arXiv",
    primaryClass = "hep-ex",
    reportNumber = "CERN-PH-EP-2012-350",
    doi = "10.1103/PhysRevD.87.052018",
    journal = "Phys. Rev. D",
    volume = "87",
    number = "5",
    pages = "052018",
    year = "2013"
}

@article{Anderle:2021hpa,
    author = "Anderle, Daniele Paolo and Dong, Xin and Hekhorn, Felix and Kelsey, Matthew and Radhakrishnan, Sooraj and Sichtermann, Ernst and Xia, Lei and Xing, Hongxi and Yuan, Feng and Zhao, Yuxiang",
    title = "{Probing gluon helicity with heavy flavor at the Electron-Ion Collider}",
    eprint = "2110.04489",
    archivePrefix = "arXiv",
    primaryClass = "hep-ex",
    reportNumber = "PhysRevD.104.114039",
    doi = "10.1103/PhysRevD.104.114039",
    journal = "Phys. Rev. D",
    volume = "104",
    number = "11",
    pages = "114039",
    year = "2021"
}

@article{HadStruc:2022yaw,
    author = "Egerer, Colin and others",
    collaboration = "HadStruc",
    title = "{Toward the determination of the gluon helicity distribution in the nucleon from lattice quantum chromodynamics}",
    eprint = "2207.08733",
    archivePrefix = "arXiv",
    primaryClass = "hep-lat",
    reportNumber = "JLAB-THY-22-3663",
    doi = "10.1103/PhysRevD.106.094511",
    journal = "Phys. Rev. D",
    volume = "106",
    number = "9",
    pages = "094511",
    year = "2022"
}

\end{document}